\documentclass[sigconf]{acmart}
\AtBeginDocument{%
  }

\usepackage{CJKutf8}
\usepackage{color}
\usepackage{multirow}
\def\ie{\textit{i.e.}~}
\def\wrt{\textit{w.r.t.}~}
\def\eg{\textit{e.g.}~}
\def\etal{\textit{et.al.}~}

\usepackage[utf8]{inputenc}
\usepackage{bbding}

\usepackage{graphicx}
\usepackage{float}
\usepackage{subfig}

\usepackage{balance}

\usepackage{stfloats}

\setcopyright{acmcopyright}
\copyrightyear{2023}
\acmYear{2023}
\setcopyright{acmlicensed}\acmConference[MM '23]{Proceedings of the 31st ACM International Conference on Multimedia}{October 29-November 3, 2023}{Ottawa, ON, Canada}
\acmBooktitle{Proceedings of the 31st ACM International Conference on Multimedia (MM '23), October 29-November 3, 2023, Ottawa, ON, Canada}
\acmPrice{15.00}
\acmDOI{10.1145/3581783.3612156}
\acmISBN{979-8-4007-0108-5/23/10}

\begin{document}

\title{ChinaOpen: A Dataset for Open-world Multimodal Learning}


\author{Aozhu Chen}
\affiliation{%
\institution{School of Information, Renmin University of China}
\city{Beijing}
\country{China}
}
\author{Ziyuan Wang}
\affiliation{%
\institution{School of Information, Renmin University of China}
\city{Beijing}
\country{China}
}
\author{Chengbo Dong}
\affiliation{%
\institution{School of Information, Renmin University of China}
\city{Beijing}
\country{China}
}
\author{Kaibin Tian}
\affiliation{%
\institution{School of Information, Renmin University of China}
\city{Beijing}
\country{China}
}
\author{Ruixiang Zhao}
\affiliation{%
\institution{School of Information, Renmin University of China}
\city{Beijing}
\country{China}
}
\author{Xun Liang}
\affiliation{%
\institution{School of Information, Renmin University of China}
\city{Beijing}
\country{China}
}
\author{Zhanhui Kang}
\affiliation{%
\institution{Tencent}
\city{Shenzhen}
\country{China}
}
\author{Xirong Li}
\authornote{Corresponding author: Xirong Li (xirong@ruc.edu.cn)}

\affiliation{%
\institution{MoE Key Lab of DEKE, Renmin University of China}
\city{Beijing}
\country{China}
}

\renewcommand{\shortauthors}{Aozhu Chen and \etal}

\begin{abstract}
This paper introduces \emph{ChinaOpen}, a dataset sourced from \emph{Bilibili}, a popular Chinese video-sharing website, for open-world multimodal learning. While the state-of-the-art multimodal learning networks have shown impressive performance in automated video annotation and cross-modal video retrieval,  their training and evaluation are primarily conducted on YouTube videos with English text. Their effectiveness on Chinese data remains to be verified. In order to support multimodal learning in the new context, we construct \emph{ChinaOpen-50k}, a webly annotated training set of 50k Bilibili videos associated with user-generated titles and tags. Both text-based and content-based data cleaning are performed to remove low-quality videos in advance. For a multi-faceted evaluation, we build \emph{ChinaOpen-1k}, a manually labeled test set of 1k videos. Each test video is accompanied with a manually checked user title and a manually written caption. Besides, each video is manually tagged to describe objects / actions / scenes shown in the visual content. The original user tags are also manually checked. Moreover, with all the Chinese text translated into English, ChinaOpen-1k is also suited for evaluating models trained on English data. In addition to ChinaOpen, we propose Generative Video-to-text Transformer (GVT) for Chinese video  captioning. We conduct an extensive evaluation of the state-of-the-art single-task / multi-task models on the new dataset, resulting in a number of novel findings and insights. 
\end{abstract}

\ccsdesc[500]{Information systems~Multimedia databases}
\ccsdesc[500]{Computing methodologies~Visual content-based indexing and retrieval}

\keywords{Chinese video dataset, multimodal learning, multi-task evaluation}

\maketitle

\section{Introduction}

\begin{figure}[htbp]
	\centering
	\subfloat[ChinaOpen-50k for training]{\label{fig:a}\includegraphics[width=2.6in]{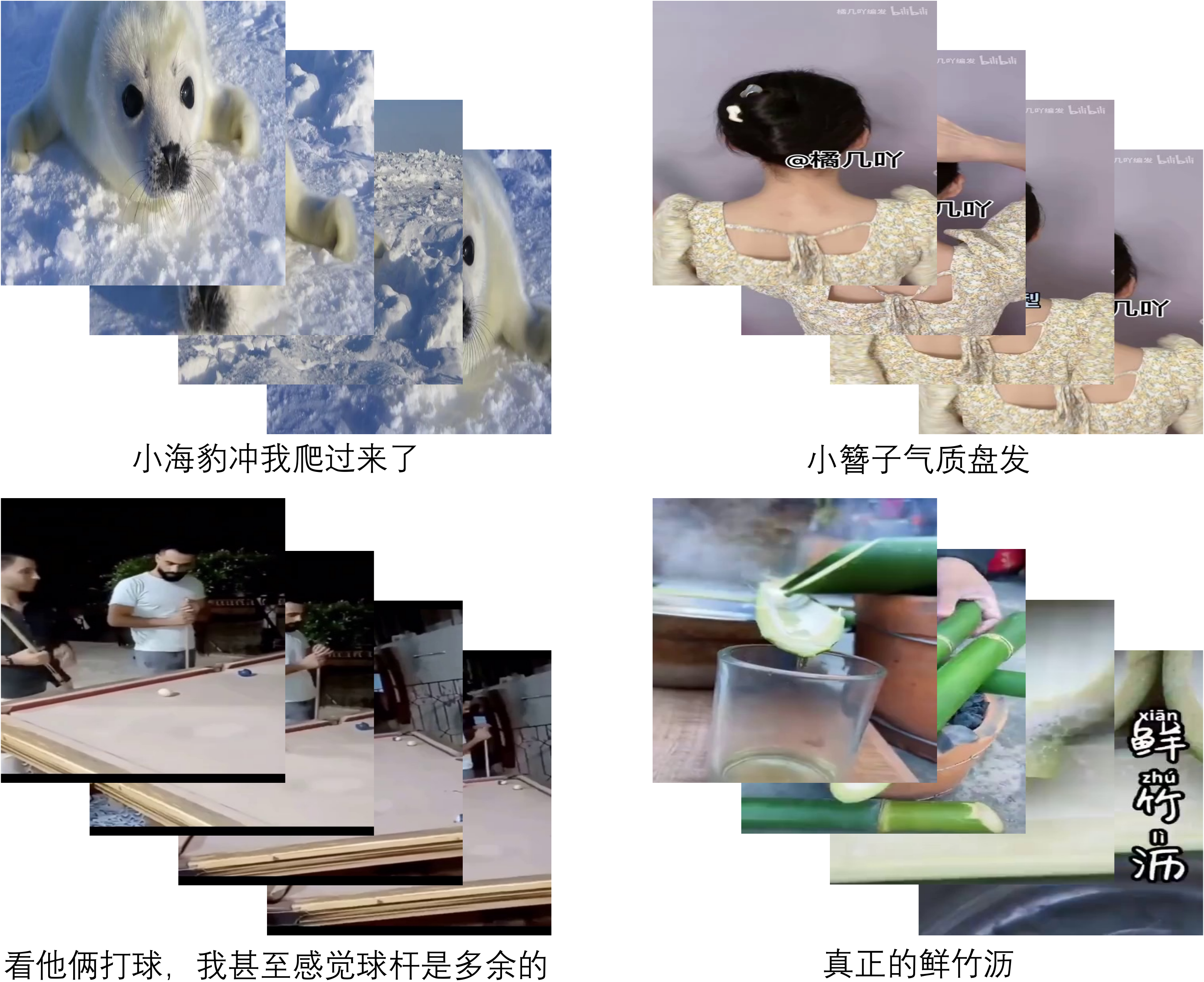}}\\
	\subfloat[ChinaOpen-1k for evaluation]{\label{fig:b}\includegraphics[width=2.6in]{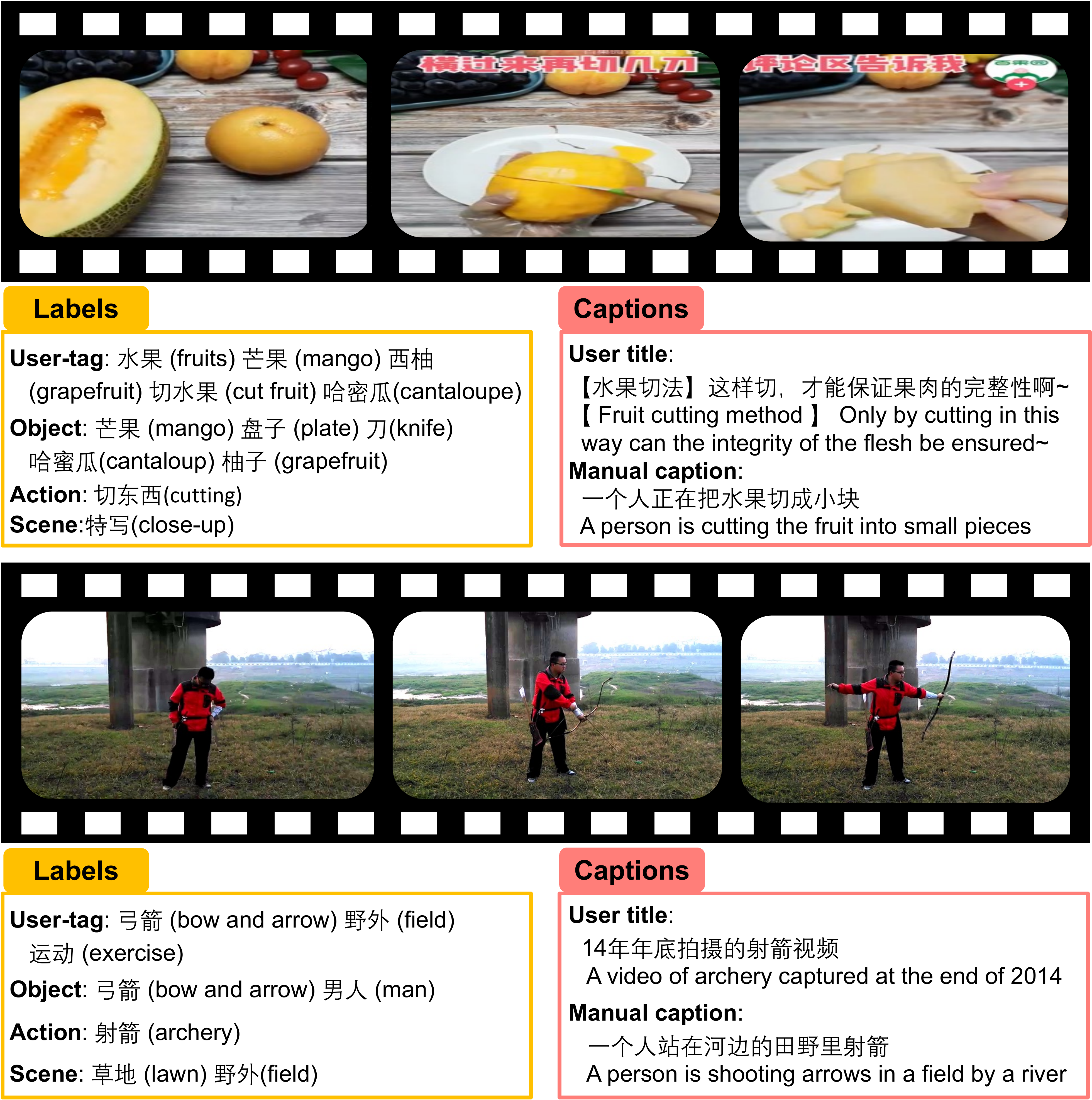}}\\	
	\caption{Visual examples and annotations from ChinaOpen: (a) ChinaOpen-50k, a selected user-titled video set  for multimodal learning and (b) ChinaOpen-1k, a manually-annotated video set for evaluating multimodal models.}
 \label{fig:visual-examples}
\end{figure}

Online short videos, typically tens of seconds to several minutes in length, are playing an increasingly important role in information dissemination on various social media platforms and video-sharing websites. Created / re-edited by individual amateurs or professional groups, these videos cover a wide range of topics from entertainment and humor to educational and informative content in a very broad domain. For visual content-based indexing and retrieval, an open world with uncontrolled content has emerged. 

While the state-of-the-art multimodal learning networks have shown impressive performance in automated video annotation \cite{git,alayrac2022flamingo} and cross-modal video retrieval \cite{clip4clip,xclip,eccv2022-laff}, their training and evaluation are primarily conducted on YouTube videos with English text. To what extent can these models generalize to Chinese data remains open. We aim to fill the gap with \emph{ChinaOpen}, a new video dataset for open-world multimodal learning and evaluation.

As exemplified in Fig. \ref{fig:visual-examples}, ChinaOpen consists of two subsets: \emph{ChinaOpen-50k} and \emph{ChinaOpen-1k}. The former is a set of 50k highly selected videos with user-generated titles (and other meta data). The latter consists of 1k videos with manually checked user-generated titles / tags, manually written captions, and manual labels describing visual objects / actions / scenes present in the
video content. 

Due to the rapidly increasing need of training large video-language models, several webly-annotated video datasets have been developed, \eg HowTo100M \cite{howto100m} and WebVid \cite{webvid}. HowTo100M consists of narrative video clips collected from YouTube, with transcribed text as their annotations. WebVid was sourced from Shutterstock with professionally edited descriptions. Both datasets were from English websites with English text. By contrast, ChinaOpen was sourced from Bilibili\footnote{\url{https://www.bilibili.com/}}, a leading video-sharing website in China, with about 90 million daily active users. ChinaOpen is thus unique.

Existing manually annotated video datasets are either label based, originally developed for video classification \cite{hvu2020,k400}, or caption based \cite{msrvtt,vatex} without manual label. ChinaOpen-1k is unique, as each video has manually labeled Chinese tags that explicitly describe objects,
actions and scenes shown in the video content. Compared to existing datasets, ChinaOpen-1k has a number of novel labels related to objects, actions and scenes, see Table \ref{tab:label-comparison}. Moreover, the video is also accompanied with a (manually checked) user-generated title and a manually written content-based caption. We argue that the title of a given video shall be sourced from its uploader as this specific user knows the context of the video and is thus in a good position to write an eye-catching title.

\begin{table}[!htb]
\caption{ChinaOpen-1k \emph{vs} public datasets in terms of their label sets. Novel labels are those unique in ChinaOpen-1k.}
\label{tab:label-comparison}
 \scalebox{0.7}{
\begin{tabular}{@{}lrrr@{}}
\toprule
\textbf{Dataset} & \textbf{Modality} & \textbf{Labels in common} & \textbf{Novel labels} \\
\midrule
\multicolumn{4}{@{}l}{\textit{\textbf{Objects}}} \\
VidOR \cite{vidor} & vid & 43 & 537 \\
MSCOCO \cite{mscoco} & img & 45 & 535 \\
Objects365 \cite{Objects365} & img & 131 & 449 \\
OpenImages \cite{OpenImages} & img & 188 & 392 \\
HVU \cite{hvu2020} & vid & 226 & 354 \\
LVIS \cite{LVIS2019} & img & 229 & 351 \\
VisualGenome \cite{visualgenome2017} & img & 231 & 349 \\
COCO-CN \cite{cococn} & img & 362 & 218 \\
\midrule
\multicolumn{4}{@{}l}{\textit{\textbf{Actions}}}  \\
UCF-101 \cite{ucf101_2012} & vid & 9 & 457 \\
Sports-1M \cite{sports1m2014} & vid & 16 & 450 \\
Kinetics-600 \cite{k600} & vid & 130 & 336 \\
Kinetics-400 \cite{k400} & vid & 139 & 327 \\
Kinetics-700 \cite{Kinetics700} & vid & 139 & 327 \\
HVU & vid & 146 & 320 \\
\midrule
\multicolumn{4}{@{}l}{\textit{\textbf{Scenes}}} \\
HVU & vid & 44 & 86\\
Places365 \cite{places365} & img & 78 & 52 \\
\bottomrule
\end{tabular}
}
\end{table}


In sum, this paper makes the following contributions: \\
$\bullet$ \textbf{Data}. We build \emph{ChinaOpen}, with ChinaOpen-50k for multimodal learning and ChinaOpen-1k for multimodal model evaluation. To the best of our knowledge, ChinaOpen is the first of its kind\footnote{ChinaOpen is available at \url{https://ruc-aimc-lab.github.io/ChinaOpen/}}. \\
$\bullet$ \textbf{Model}. We propose Generative Video-to-text Transformer (GVT) for Chinese video captioning. GVT improves over GIT \cite{git} with a simple visual-token reduction layer that effectively scales up the number of input video frames, resulting in  better performance. \\
$\bullet$ \textbf{Evaluation}. We evaluate up to 15 SOTA models  (11 English and 4 Chinese), see Table \ref{tab:sotas}. Our evaluation covers up-to-date developments, \eg  ERNIE-ViL2 \cite{ernievil2}, CN-CLIP \cite{cnclip} and Taiyi \cite{taiyi} for open-set video tagging, X-CLIP \cite{xclip} for text-to-video retrieval, Flamingo \cite{alayrac2022flamingo}, GIT, mPLUG \cite{mplug} and BLIP-2 \cite{blip2} for video captioning.

\begin{table}[b!]
\caption{SOTA models evaluated on ChinaOpen-1k.}
\label{tab:sotas}
 \scalebox{0.7}{
 \begin{tabular}{@{} lrlcccc @{}}
\toprule
\textbf{Model} & \textbf{\#params} & \textbf{Vision} & \textbf{Lang.} & \textbf{Tagging} & \textbf{Retrieval} & \textbf{Captioning} \\
\hline
ResNet-P365 \cite{places365} & 24M & img & EN & \Checkmark & \XSolidBrush & \XSolidBrush \\
SwinB-K400 \cite{k400-swint} & 88M & vid & EN & \Checkmark & \XSolidBrush & \XSolidBrush \\ 
\midrule

CLIP4Clip \cite{clip4clip} & 164M & vid & EN & \XSolidBrush & \Checkmark & \XSolidBrush \\
X-CLIP \cite{xclip} & 164M & vid & EN & \XSolidBrush & \Checkmark & \XSolidBrush \\ 
\midrule

CLIP-32/B \cite{clip} & 151M & img & EN & \Checkmark & \Checkmark & \XSolidBrush \\
CN-CLIP \cite{cnclip} & 188M & img & CN & \Checkmark & \Checkmark & \XSolidBrush \\
ERNIE-ViL2 \cite{ernievil2}  & 204M & img & CN & \Checkmark & \Checkmark & \XSolidBrush \\
Taiyi \cite{taiyi} & 254M & img & CN & \Checkmark & \Checkmark & \XSolidBrush \\
CLIP-L/14@336px \cite{clip} & 427M & img & EN & \Checkmark & \Checkmark & \XSolidBrush \\  \midrule
OFA-Chinese \cite{wang2022ofa} & 160M & img & CN & \XSolidBrush & \XSolidBrush & \Checkmark \\
GIT \cite{git} & 161M & img/vid & EN & \XSolidBrush & \XSolidBrush & \Checkmark \\
BLIP \cite{li2022blip} & 247M & img & EN & \XSolidBrush & \XSolidBrush & \Checkmark \\
mPLUG \cite{mplug} & 574M & img & EN & \XSolidBrush & \XSolidBrush & \Checkmark \\
Flamingo \cite{alayrac2022flamingo} & 1,138M & img/vid & EN & \XSolidBrush & \XSolidBrush & \Checkmark \\
BLIP-2 \cite{blip2} & 3,745M & img & EN & \XSolidBrush & \XSolidBrush & \Checkmark\\
\midrule
GVT (\emph{this paper}) & 146M & vid & CN & \XSolidBrush & \XSolidBrush & \Checkmark\\
\bottomrule
\end{tabular}
}
\end{table}

The rest of the paper is organized as follows. We discuss related work in Sec. \ref{sec:related}. ChinaOpen is detailed in Sec. \ref{sec:dataset}, followed by GVT in Sec. \ref{sec:gvt} and evaluation in Sec. \ref{sec:eval}. Conclusions are given in Sec. \ref{sec:conc}.

\section{Related Work}\label{sec:related}
Webly-annotated video data is crucial for developing large multimodal learning models. Meanwhile, manually-annotated video data is a must for properly evaluating the developed models. We briefly review progress in these two subjects, explaining accordingly 
 how ChinaOpen uniquely contributes to the field.

\textbf{Progress on webly-annotated video datasets}. 
Due to the growing need of training large multimodal models for video-language related tasks, there have been good efforts on harvesting weakly-annotated videos from the web \cite{howto100m,webvid}. HowTo100M \cite{howto100m}, consisting of hundred million narrated video clips, has been used for learning text-video embedding. Despite the extremely large scale, the transcribed texts from narrated video clips are over noisy that specifically designed noise-tolerant learning algorithms have to be used. A more recent dataset, WebVid, contains 10M short videos with their textual descriptions sourced from stock footage sites \cite{webvid}. In particular, the authors of \cite{webvid} have used a 2.5M subset of WebVid to train a deep video-text matching model. 
We observe that these descriptions tend to be carefully worded to describe the video content in a concise manner, making them differ substantially from common user-generated titles. Also note that both HowTo100M and WebVid were sourced from English websites. In this context, our ChinaOpen-50k, which consists of user-titled videos from a popular Chinese video-sharing website, is unique. Moreover, we develop automated data cleaning to exclude videos that are either with low-quality annotations or lacking meaningful visual elements. As such, ChinaOpen-50k, while being relatively small-scale, already  shows a good potential in our experiments.

\textbf{Progress on manually-annotated video datasets}. Existing manually-annotated video datasets mostly focus on a specific task, including human action recognition (HMDB \cite{hmdb2011} and UCF-101 \cite{ucf101_2012}), sports-related activities (Sports-1M  \cite{sports1m2014}) or a broader range of actions (Kinetics-400 \cite{k400}). The Holistic Video Understanding (HVU) dataset expands the labels from actions to scenes, objects, attributes and concepts \cite{hvu2020}. However, the annotations of the above datasets are in the form of labels, making them unsuited for video-language tasks such as text-to-video retrieval which matches videos and natural-language text and video captioning that generates textual descriptions of the video content. Meanwhile, current video-text datasets such as MSVD \cite{msvd}, MSR-VTT \cite{msrvtt}, and VaTeX \cite{vatex} has no manual label. Our ChinaOpen-1k dataset is unique as each video has manually labeled Chinese tags that explicitly describe objects, actions and scenes shown in the video content. Moreover, the video is also accompanied with a (manually checked) user-generated title and a manually-written content-based caption.

\section{The $ChinaOpen$ Dataset} \label{sec:dataset}

\begin{figure*}[htbp]
    \centering
    \includegraphics[width=2\columnwidth]{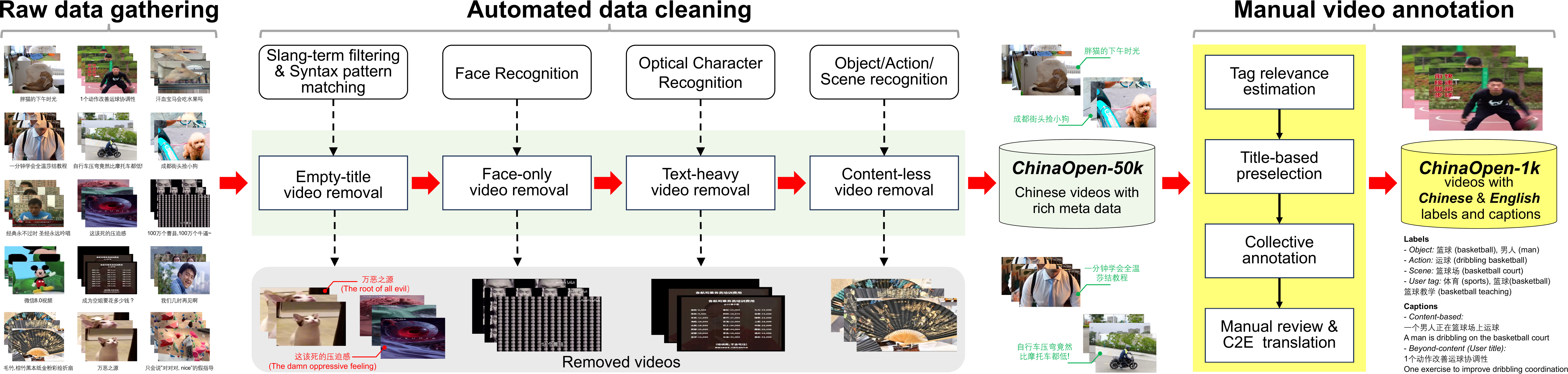}
    \caption{Conceptual diagram of the construction of the proposed \emph{ChinaOpen} dataset. Given a set of 100k \emph{Chinese} videos randomly gathered from Bilibili (Sec. \ref{ssec:data-gather}), we perform automated data cleaning to remove videos either with low-quality annotations or lacking meaningful visual content (Sec. \ref{ssec:data-cleaning}). This leads to \emph{ChinaOpen-50k}, a webly-labeled set of 50k videos for multimodal learning. For a multi-faceted evaluation, we build a ground-truthed test set \emph{ChinaOpen-1k} (Sec. \ref{ssec:data-annotation}). Each test video is accompanied with a manually-checked user title, a manually-written content-based caption, manually-checked user tags, and a number of labels describing visual objects / actions / scenes present in the video content. With all the Chinese text translated to English, ChinaOpen-1k is also suited for directly evaluating multimodal models trained on English data. }
    \label{fig:ChinaOpen}
\end{figure*}

As Fig. \ref{fig:ChinaOpen} shows, the ChinaOpen dataset is constructed in three stages. That is, raw data gathering from Bilibili, automated data cleaning to obtain ChinaOpen-50k (for multimodal learning), and lastly manual video annotation to produce ChinaOpen-1k (for multi-task evaluation). We depict each stage in the following.

\subsection{Raw Data Gathering}\label{ssec:data-gather}

In order to obtain a representative and diverse subset of Bilibili short videos, we randomly generated many video ids as candidates. For this study, we gathered nearly 100k videos uploaded between May 2010 and Sep. 2021. Besides the MP4 video files, we also downloaded varied meta data, including titles, tags, descriptions, comments and danmaku (a.k.a. flying comments), if available. The duration of the downloaded videos ranges from 2 seconds to 608 seconds, with a mean value of 28.4 seconds and median of 25. The number of Chinese characters per video title ranges from 1 to 80, with a mean value of 16.4 and median of 14. The Bilibili platform organizes user-uploaded videos in channels, which are Bilibili-defined keywords that describe the videos at a very high level. Videos in ChinaOpen were collected from nearly 100 channels, where the top 10 channels are \textit{daily}, \textit{funny}, \textit{celebrity}, \textit{society}, \textit{film} and \textit{TV editing}, \textit{general}, \textit{beauty and skincare}, \textit{body building}, \textit{cat}, and \textit{outfit}.


\subsection{Automated Data Cleaning} \label{ssec:data-cleaning}
 
As the raw data is quite diverse with varied annotation quality, automated data cleaning is necessary to remove videos which are either with low-quaity annotations or lacking meaningful visual content for a broad audience. With manual screening on the raw data, we empirically find that unwanted videos can be largely attributed to the following four categories, \ie \emph{empty-title}, \emph{face-only}, \textit{text-heavy}, and \emph{content-less}. To that end, we develop as follows a multimodal method to identify videos of the four categories step-by-step, and remove them accordingly.

\subsubsection{Empty-title video removal}

We consider a video title empty if it has no verb-noun phrase (VNP) and thus tells little about the video content. In order to determine the presence of VNPs, we parse the given title with HanLP \cite{hanlp}, an open-source Chinese NLP toolbox\footnote{\url{https://github.com/hankcs/HanLP}}, to obtain a syntactical representation of the title. Accordingly, syntax pattern matching is performed to find verb-object structures, nominal phrases with modifier-head constructions, and subject-verb-object constructions within the representation. 

\begin{CJK*}{UTF8}{gbsn}
Note that due to the Chinese video-sharing culture, many titles are mainly  comprised of slang terms such as ``名场面'' (iconic scene), ``打卡挑战'' (daily attendance), and ``跟着UP主创作吧'' (follow the uploader to create it). These terms are so frequently used that they tell little about the video content, and thus form the basis of our stopword list. We further expand the list to cover mental verbs such as ``觉得'' (think), ``知道'' (know) and ``建议'' (suggest) and non-Chinese characters such as punctuation, emoticons and Japanese / Korean characters. The title of a given video is filtered with the stopword list, followed by syntax pattern matching. If no VNP is found, the given video will be removed.
\end{CJK*}

\subsubsection{Face-only video removal}

Face-only videos show mostly faces, with little background or other visual elements. More specifically, we observe two major patterns, \ie \emph{talking head} and \emph{face mosaic}. A talking-head video shows only a person's head (and shoulders), while a face-mosaic video contains frames showing a collage of many (small) faces. We therefore resort to frame-wise face detection. We adopt InsightFace\footnote{\url{https://insightface.ai/}}, an open-source deep face analysis library, with its default detection model (SCRFD-10GF). For a given video, we sample its frames uniformly. 
Face detection is performed per frame. A frame is classified as talking-head if a detected face region is over 50\% of the image area. Accordingly, the given video talking-head is treated as talking-head if more than 75\% of its frames are talking-head. To determine if the video is face-mosaic, we count the maximum number of faces detected per frame. If the number exceeds a given threshold (which is empirically set to 8), the video is labeled as face-mosaic. As such, we remove face-only videos.

\subsubsection{Text-heavy video removal}

Video with many texts on their frames typically lack visual elements of common interest. In order to filter out such text-heavy videos, we conduct Optical Character Recognition (OCR) on frames. In particular, we adopt PaddleOCR\footnote{\url{https://github.com/PaddlePaddle/PaddleOCR}}, a public and leading OCT tool that recognizes multilingual texts from images. A frame is considered as text-heavy, if the number of OCR-detected characters exceeds 50. Similarly, we consider a video text-heavy if more than 75\% of its frames are text-heavy.

\subsubsection{Content-less video removal}

We consider a given video content-less if it lacks recognizable object, action or scene. To that end, we employ existing visual recognition models to estimate if there is any object / action / scene present in the given video. For scene recognition, we employ a ResNet-152 network trained on the Places-365 image dataset \cite{places365} (ResNet-P365). As the input of ResNet-P365 shall be an image, we simply take the middle frame of the given video. For action recognition, we utilize a Video Swin Transformer (Swin-B as its backbone) \cite{k400-swint} trained on the Kinetics-400 video dataset \cite{k400}, which we term SwinB-K400. Note that the 365 scene / 400 action classes defined in Places365 / Kinetics-400 are insufficient to cover the rich content of the Chinese videos. ResNet-P365 tends to incorrectly categorize videos of dogs or cats as ``veterinarians office'', while SwinB-K400 tends to mistakenly label videos of cooking as ``cooking chicken''. Despite such biases, their predictions remain instructive to filter out content-less videos.

For object recognition, we curate a set of 3,841 object labels by merging object classes from  MSCOCO \cite{mscoco}, VisualGenome \cite{visualgenome2017}, Objects365 \cite{Objects365}, LVIS \cite{LVIS2019} and OpenImages \cite{OpenImages}. In order to predict the relevance of these objects \wrt the video, we use a pre-trained CLIP model (ViT-B/32) \cite{clip} for zero-shot tagging on the middle frame. 
With the English labels manually translated to Chinese, we further employ CN-CLIP \cite{cnclip}, a Chinese version of CLIP, to tag the video with the Chinese object labels.

Object-wise, we consider a video not content-less if the video is predicted with at least one highly confident label (cutoff at the 75th percentile) or two moderately confident labels (cutoff at the 50th percentile). Action and scenes labels are postprocessed in a similar manner. A video is treated as content-less if no label is emitted.

With the video cleaning process described above, we obtain a cleaned set of 50k videos, termed ChinaOpen-50k. Video duration is between 3 seconds to 608 seconds, with a mean value of 29.8 and median of 27. File size per video is between 81.9KB and 20.9MB, with a mean value of 1.4MB and a median value of 1.2MB. ChinaOpen-50k has 431.2 hours and 69.1 GB of videos in total. With 19.2 characters on average, video titles are longer than those in the raw data (16.4 characters on average). More importantly, compared to the raw data, ChinaOpen-50k provides a better starting point for both multimodal learning and fine-grained manual annotation.

In addition to the user-generated titles, we enrich the annotations of ChinaOpen-50k by auto captioning.  We adopt an existing model  \cite{DongCCHW021}, trained on a joint set of  MSR-VTT \cite{msrvtt}, VaTeX \cite{vatex}, TGIF \cite{tgif2015} and Action-GIF \cite{autogif}.  As the generated captions are in English, we use machine translation\footnote{\url{https://fanyi-api.baidu.com/}} to convert them to Chinese.

\subsection{Manual Video Annotation}\label{ssec:data-annotation}

As aforementioned, we aim to build a ground-truthed Chinese video dataset to support multi-task evaluation. The tasks include general-purpose video content recognition (objects, actions, and scenes), assisted user tagging / captioning, and video retrieval by natural-language text. Since manual annotation is known to be expensive and thus much limited, video preselection is necessary to make the manual annotation process well pay off.

\subsubsection{Video preselection}

User tags are known to be subjective and personalized \cite{LiITM2009}. In order to find from ChinaOpen-50k videos that are likely to be accompanied with content-relevant tags, we adopt the classical neighbor voting algorithm \cite{LiITM2009}. Per video, we retrieve its 200 neighbors from the dataset in terms of cosine similarity between the video-level CLIP features. A user tag associated with the query video is deemed to be visually relevant if the tag  also appears in the user-tag list of the neighbor videos. Next, from the videos with at least one content-related tag, we randomly sample 10k videos for title-based preselection as follows. 

In contrast to a machine-generated caption trying to objectively describe what is visible, a user-generated title tends to be more eye-catching, providing readers with a beyond-content interpretation of the video. To strike a proper balance between relevance and attractiveness, we prefer to choosing videos with relevant titles such that a common user can easily relate the titles to the video content. Following this criterion, a review board of three experienced annotators read the titles of the 10k sampled videos, accordingly selecting 3k videos for collective annotation. To reduce the annotation workload, videos exceeding 60 seconds are excluded beforehand.

\subsubsection{Collective annotation}

Our annotation team consists of 16 members who are staffs and students in our lab. Each annotator has been instructed to annotate a given video in a coarse-to-fine manner. Firstly, the annotator is asked to check again if the user title is indeed content-relevant. If the answer is negative, the video will be skipped. Second, the annotator writes a caption that shall faithfully describe the gist of the video content. Next, the annotator describes with Chinese labels what objects / actions / scenes are shown in the video. Lastly, the annotator checks if the user-provided tags are content-relevant. To ensure the annotation richness, a video with zero label in a specific aspect (objects, actions, scenes or user tags) will be discarded.

\subsubsection{Manual review}
\begin{CJK*}{UTF8}{gbsn}
After the collective annotation stage, the review board performs a double check on the annotations for two purposes. That is, to fix labeling issues occasionally made by individual annotators and to translate the Chinese captions and labels to English assisted by machine translation. In total, we have 1,092 videos manually annotated with 1,092 user-generated titles, 1,092 content-based captions, 7,910 Chinese tags and 7,856 English tags\footnote{Chinese tags such as ``喵星人'', ``猫'', ``猫咪'' are  translated to ``cat'', so the number of English tags is relatively smaller.} in total. 
\end{CJK*}
The number of distinct Chinese / English tags is 2,100 / 2,030.
Compared to existing (Chinese) video captioning datasets, \eg VaTeX-CN, which have content-based captions only, the availability of user titles allows us to evaluate models in a novel beyond-content track. While targeted at Chinese models, the availability of  English annotations also allows us to evaluate English models. We term the new testset ChinaOpen-1k. 

The testset has 9.9 hours and 1.6 GB of videos in total. Playback duration per video is between 5 seconds to 60 seconds, with a mean value of 32.5 seconds and a median value of 30 seconds. File size is between 0.2MB to 3.4MB, with mean value of 1.5MB and median value of 1.4MB. The size of the Chinese object / action / scene / user-tag vocabulary  is 580 / 466 / 130 / 924. Table \ref{tab:table-annotation-statistics} shows annotation statistics in details. Compared with existing label-based datasets, ChinaOpen-1k has hundreds of unique labels, see Table \ref{tab:label-comparison}. There is no overlap between ChinaOpen-50k and ChinaOpen-1k videos.

\begin{table}[htbp!]
\caption{\textbf{Annotation statistics of ChinaOpen-1K}.}
\label{tab:table-annotation-statistics}
\centering
 \scalebox{0.7}{
 \begin{tabular}{@{} lrrrr @{}}
\toprule
\textbf{Annotations}  & \textbf{Min} & \textbf{Max} & \textbf{Mean} & \textbf{Median}\\ \hline
\multicolumn{5}{@{}l}{\textit{Number of Chinese labels per video:}} \\ 
Objects &1 &9 &2.45 &2 \\ 
Actions &1 &6 &1.37 &1 \\ 
Scenes &1 &3 &1.10 &1 \\ 
Verified user-tags &1 &9 &2.33 &2  \\ \midrule

\multicolumn{5}{@{}l}{\textit{Number of characters per Chinese caption}:} \\ 
User-generated title &6 &79 &18.88 &17 \\ 
Content-based caption &5 &38 &14.23 &14 \\ \bottomrule

\end{tabular}
}
\end{table}

\section{Multimodal Learning on ChinaOpen} \label{sec:gvt}

As a showcase of multimodal learning on ChinaOpen-50k, we describe in this section how to train a Transformer-based Chinese video captioning model on the webly-annotated dataset. 

We depart from the Generative Image-to-text Transformer (GIT) \cite{git}, a state-of-the-art model on multiple vision-to-language generation tasks including image captioning, video captioning, and VQA. Consider image captioning for instance. At the training stage, given an input image of size $256 \times 256$ and a reference caption of $m$ words, GIT uses a Vision Transformer (ViT) to encode the input image, generating an array of $14\times 14$=196 visual tokens plus a special [CLS] token, each with a 768-d embedding vector. Meanwhile, the caption is also tokenized and represented by an array of $(m+2)\time 768$ embeddings, where the two extra tokens indicate the beginning and the end of the sentence, a.k.a. [BOS] and [EOS]. The visual and textual tokens are then concatenated and fed into a language Transformer for text generation. Note that in order to prevent information leakage during decoding, causal self-attention is used in the language Transformer such that each textual token is only permitted to ``see'' its preceding tokens, \ie all visual tokens and the textual tokens before the current token. For video captioning, GIT simply concatenates the visual tokens of the input video frames. Given $k$ frames as input, such a strategy will yield a large number of $k\times197$ visual tokens. Consequently, GIT has to set $k$ to be a small number (which is 6) to make the computation feasible. However, the small $k$ means sparse sampling of the video frames, inevitably causing much loss in the visual information.

\begin{figure}[htb!]
    \centering
    \includegraphics[width=0.97\columnwidth]{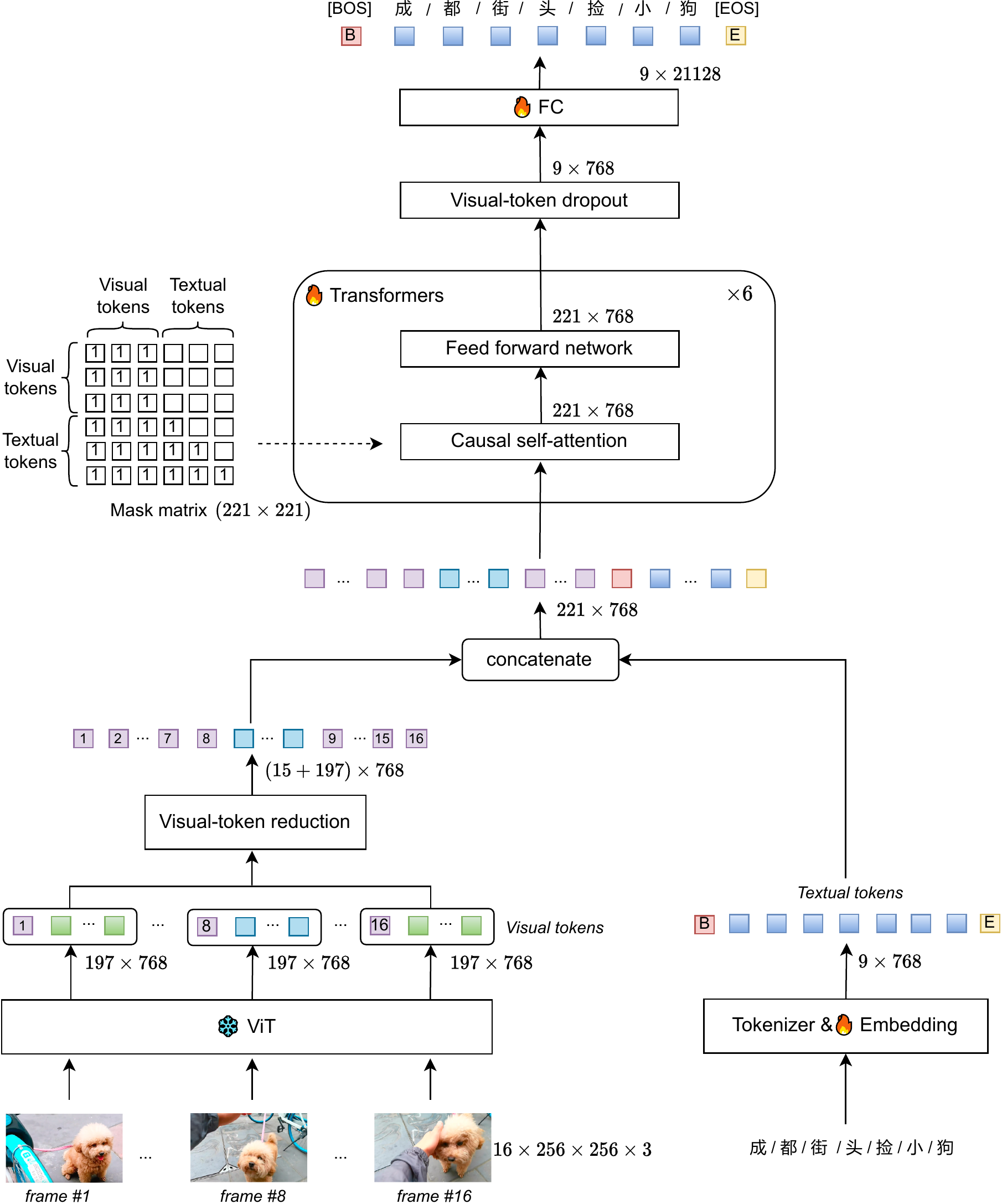}
    \caption{Proposed Generative Video-to-text Transformer (GVT) for video captioning. GVT improves over GIT with a simple \emph{visual-token reduction} layer that effectively scales up the number of input video frames, from 6 in GIT to 16.  }
    \label{fig:gvt}
\end{figure}

 Note that for a specific frame, its [CLS] token has been updated by the other tokens of the frame within ViT. Hence, the [CLS] token represents the frame to a large extent. Also note that for a given video, its middle frame is typically more representative than the other frames.  Hence, selectively combining all the tokens from the middle frame and the [CLS] tokens of the other frames seems reasonable. We implement this idea with a simple visual-token reduction (VTR) layer, see Fig. \ref{fig:gvt}. With VTR, the number of visual tokens to be fed into the language Transformer is substantially reduced from $k\times 197$ to $k+196$. Such a minor tweak\footnote{We also try adding all tokens from the first and the last frames. Accordingly, the number of visual tokens increases from $k$+196 to $k$+588. Adding more frames marginally improves the performance, yet with noticeably increased computational overhead.} allows us to effectively scale up the number of input video frames from 6 to 16. With the VTR layer, GIT is tailored to video captioning. We term the improved model Generative Video-to-text Transformer (GVT). 
 
The vision and language Transformers of GVT are initialized by a pre-trained GIT\_BASE\footnote{\url{https://publicgit.blob.core.windows.net/data/output/GIT_BASE/snapshot/model.pt}}. In order to cope with Chinese, the language-specific layers, \eg text tokenizer, token embedding and the last FC layer, are re-trained from scratch if applicable.

\section{Evaluation} \label{sec:eval}

\subsection{Common Setup}

Subject to the availability of a model's PyTorch training / inference code and our computation capacity (8$\times$NVIDIA RTX 3090 GPUs), we collect the following SOTA models: \\
$\bullet$ \textbf{Video tagging} (2): \emph{ResNet-P365}  (ResNet-152 trained on Places365) \cite{places365} and \emph{SwinB-K400} \cite{k400-swint} (Video Swin Transformer with Swin-B as its backbone trained on the Kinetics-400 video action dataset.  \\
$\bullet$ \textbf{Text-to-video retrieval} (2): \emph{CLIP4Clip} \cite{clip4clip} (Transferring CLIP to the video domain with a sequential Transformer for temporal modeling) and \emph{X-CLIP} \cite{xclip} (improving over CLIP4Clip with fine-grained cross-modal matching). \\
$\bullet$ \textbf{Tagging \& retrieval} (5): \emph{CLIP-B/32} \cite{clip} (ViT-B/32 as visual encoder and GPT-2 as text encoder), \emph{CLIP-L/14@336px} \cite{clip} (ViT-L/14-336px as visual encoder and GPT-2 as text encoder), \emph{Taiyi} \cite{taiyi} (ViT-B/32 as visual encoder and Chinese-Roberta-wwm-base as text encoder), \emph{CN-CLIP} \cite{cnclip} (ViT-B/16 as its visual encoder and Chinese-Roberta-wwm-base as text encoder), and \emph{ERNIE-ViL2} \cite{ernievil2} (ViT-B/16 as visual encoder and ERNIE-3.0-base as text encoder). \\
$\bullet$ \textbf{Video captioning} (6): \emph{OFA-Chinese} \cite{wang2022ofa} (ResNet-101 as visual encoder and Transformer as text encoder), \emph{GIT} \cite{git} (ViT-B/16 as visual encoder), \emph{BLIP} \cite{li2022blip} (ViT-B/16 as visual encoder and BERT-base as text encoder), \emph{BLIP-2} \cite{blip2} (ViT-L/14 as visual encoder and Query Transformer as text encoder), \emph{mPLUG} \cite{mplug} (ViT-B/16 as visual encoder and  6-layer Transformer as text encoder) and \emph{Flamingo} \cite{alayrac2022flamingo} (ViT-L/14 as visual encoder and OPT-350m as text encoder).


By default, each model is evaluated using the ground truth of its own language, unless stated otherwise.

\subsection{Task I: Open-Set Video Tagging}

\subsubsection{Task setup}

In open-set video tagging, a model is asked to tag a given video with an ad-hoc vocabulary that the model is not specifically tuned for. Recall that ChinaOpen-1k has tags along four dimensions, \ie objects, actions, scenes and user tags. Evaluating video tagging per dimension reveals how good the model is at recognizing objects / actions / scenes and assisting user tagging. 

By prompt-based label embedding, large multimodal models such as CLIP, CN-CLIP, Taiyi and ERNIE-ViL2 are naturally applicable for the open-set setting. However, the setting is challenging for ResNet-P365 and SwinB-K400 trained with a fixed vocabulary. Note for instance ChinaOpen-1k has 52 novel scene labels compared to Places365, see Table \ref{tab:label-comparison}. To resolve the label mismatch, we convert the prediction of ResNet-P365 as follows. Per test video, we use ResNet-P365 to predict top-5 labels for the middle frame. For each (English) tag in the ChineseOpen-1k scene vocabulary, its relevance score to the given video is calculated by summing up the scores of the predicted labels weighed by their BERT similarity scores to the tag. In a similar vein we handle SwinB-K400.

Among the three pre-trained Chinese models (CN-CLIP, ERNIE-ViL2, and Taiyi), CN-CLIP is the only model that has PyTorch training code released. We thus choose to fine-tune this model with ChinaOpen-50k and a learning rate of 1e-5. Since CN-CLIP is an image model, we simply take the middle frame per video. We also try the same fine-tuning strategy with VaTeX-CN (VaTeX with Chinese captions). For the ease of reference, we use the notation \$\{model\} [\$\{\emph{dataset}\}] to indicate \$\{model\} trained on on \$\{dataset\}.

\subsubsection{Performance metric}
We compute Average Precision (AP) per test image, as commonly used to evaluate multi-label classification \cite{csur2016-tagsurvey}. The overall performance is measured by mean AP.

\subsubsection{Results}

The performance of different models is reported in Table \ref{tab:table-tagging-AP}. Among the four English models, the CLIP series clearly surpass ResNet-P365 and SwinB-K400, showing the superiority of the large multimodal models in the open-set tagging scenario. Meanwhile, CLIP-L/14@336px is noticeably better than CLIP-B/32. Note the main difference between the two models that the former is equipped with a much larger ViT. Similar results are also observed among the Chinese models, where CN-CLIP which uses ViT-B/16 outperforms Taiyi which uses the smaller ViT-B/32. CN-CLIP fine-tuned on ChinaOpen-50k is better than the original.

\begin{table}[htbp!]
\renewcommand{\arraystretch}{1}
\caption{\textbf{Performance of open-set video tagging}. Metric: AP.}
\label{tab:table-tagging-AP}
\centering
 \scalebox{0.7}{
 \begin{tabular}{@{} lrrrrr @{}}
\toprule
\textbf{Model}  & \multicolumn{1}{l}{\textbf{Objects}} & \multicolumn{1}{l}{\textbf{Actions}} & \multicolumn{1}{l}{\textbf{Scenes}} & \multicolumn{1}{l}{\textbf{User-tags}} & \textbf{Mean} \\ \hline
\multicolumn{6}{@{}l}{\textit{English models:}}  \\ 
SwinB-K400 & - & 7.7 & - & - & - \\
ResNet-P365  & - & - & 8.7 & - & - \\
CLIP-B/32  & 34.6 & 32.8 & 32.4 & 27.9 & 31.9 \\
CLIP-L/14@336px  & \textbf{44.0} & \textbf{42.7} & \textbf{36.7} & \textbf{35.7} & \textbf{39.8} \\
\hline

\multicolumn{6}{@{}l}{\textit{Chinese models:}}\\ 
Taiyi & 38.8 & 38.2 & 45.7 & 35.7 & 39.6 \\
ERNIE-ViL2  & 40.8 & 40.0 & 47.4 & 35.3 & 40.9 \\
CN-CLIP [\emph{VaTeX-CN}] & 40.7   & 41.7  & \textbf{48.3}    & 34.4 & 41.3  \\
CN-CLIP & 39.2 & \textbf{43.0} & 47.1 & 36.4 & 41.4\\
CN-CLIP [\emph{ChinaOpen-50k}] & \textbf{42.6}  & 42.2    & 45.8  & \textbf{37.3}   & \textbf{42.0} \\

\bottomrule

\end{tabular}
}
\end{table}




Along the four dimensions, the result suggests that the top English model (CLIP-L/14@336px) recognizes objects the best, while the three Chinese models are relatively consistent, all recognizing scenes the best. For both English and Chinese models, their performance in the user-tag dimension is relatively the worst, suggesting that assisted user tagging is more challenging.

\subsection{Task II: Text-to-Video Retrieval}

\subsubsection{Task setup}

Text-to-video retrieval is to rank videos in terms of their cross-modal similarity to a given textual query. Recall that each test video is associated with a user-generated title and a manually-written caption. This allows us to setup two evaluation tracks: a content-based track which uses the manual captions as test queries and a beyond-content track which uses the user titles as test queries. 

For the large image-text models (CLIPs, CN-CLIP, Taiyi and ERNIE-ViL2), their video-level feature is obtained by mean pooling over the corresponding frame-level features. 
As for X-CLIP and CLIP4Clip originally developed for text-to-video retrieval, we use ViT-B/32 as their visual encoder and have them trained on four popular (English) video-text datasets, \ie  MSVD \cite{msvd}, MSR-VTT \cite{msrvtt}, VaTeX \cite{vatex} and ActivityNet-Caption \cite{activitynetcap}, respectively.

\subsubsection{Evaluation criteria}

We report the commonly used Recall at Rank N (R@N, N=1, 5, 10) and their summation, denoted as SumR.

\subsubsection{Results}

Text-to-video retrieval performance of the individual models are summarized in Table \ref{tab:Text-to-Video_Retrieval}. Among the English models, CLIP-L/14@336px is again the best, showing the importance of using a larger ViT. Nevertheless, the performance gap between CLIP-L/14@336px and CLIP-B/32 (233.2 versus 210.3 in SumR in the content-based track) can be effectively reduced by training a task-specific network on many video-text pairs, see X-CLIP [\emph{VaTeX}] with SumR of 229.7. The superior performance of the English models as compared to their Chinese counterparts is largely due to the use of much larger ViT, see CLIP-L/14@366px. Given model size (\#parameters) at the same level, \textit{c.f.} Table \ref{tab:sotas}, CN-CLIP (with 188M parameters) and CLIP B/32 (with 151M parameters) are largely comparable (SumR 213.2 vs 210.3). The performance of CN-CLIP is improved by fine-tuning on ChinaOpen-50k.

Comparing the two tracks, we observe that the performance gain of X-CLIP [\emph{VaTeX}] over CLIP-B/32 in the beyond-content track is much less than its counterpart in the content-based track (7.4 \emph{versus} 19.4 in SumR). The result indicates a clear discrepancy between manually-written captions and user-generated titles. 
For all models, their performance in the beyond-content track is consistently lower than in the content-based track. We conclude from the result that querying by user-titles is more difficult.

Comparing the three Chinese models, while CN-CLIP is the best for the video tagging task, ERNIE-ViL2 now outperforms CN-CLIP and Taiyi. Given that ERNIE-ViL2 and CN-CLIP use the same visual encoder (ViT-B/16) but different text encoders (ERNIE-3.0 \emph{versus} Chinese-Roberta-wwm), the result suggests that ERNIE-3.0 provides a better textual-query representation.

\begin{table}[!htb]
\caption{Performance of text-to-video retrieval. 
Models per language are sorted by their overall performance.}
\label{tab:Text-to-Video_Retrieval}
 \scalebox{0.7}{
\begin{tabular}{@{}llrrrrrrrr@{}}
\toprule
\multicolumn{1}{@{}l}{\multirow{2}{*}{\textbf{Model}}} & \multicolumn{4}{c}{\textbf{Content-based track}} & \multicolumn{1}{c}{} & \multicolumn{4}{c}{\textbf{Beyond-content track}}  \\
\cmidrule{2-5}  \cmidrule{7-10}
\multicolumn{1}{c}{}& \multicolumn{1}{l}{R@1} & \multicolumn{1}{l}{R@5} & \multicolumn{1}{l}{R@10} & \multicolumn{1}{l}{SumR} & \multicolumn{1}{l}{} & \multicolumn{1}{l}{R@1} & \multicolumn{1}{l}{R@5} & \multicolumn{1}{l}{R@10} & \multicolumn{1}{l}{SumR}  \\
\hline
\multicolumn{10}{@{}l}{\textbf{\textit{English models:}}}  \\

CLIP-B/32 & 49.5 & 75.9 & 84.9 & 210.3 &  & 32.3 & 59.0 & 68.3 & 159.6 \\
X-CLIP [\emph{MSR-VTT}] & 50.7 & 79.9 & 86.9 & 217.5 &  & 32.5 & 58.8 & 69.9 & 161.2 \\
CLIP4CLIP [\emph{MSVD}] & 51.3 & 77.3 & 86.6 & 215.2 &  & 34.6 & 60.2 & 70.6 & 165.4 \\
CLIP4CLIP [\emph{ActivityNet}] & 53.5 & 79.9 & 87.7 & 221.1 &  & 34.1 & 58.3 & 68.8 & 161.2 \\
CLIP4CLIP [\emph{MSR-VTT}] & 52.3 & 79.9 & 88.5 & 220.7 &  & 33.9 & 59.9 & 68.9 & 162.7 \\
X-CLIP [\emph{MSVD}] & 53.7 & 79.3 & 86.6 & 219.6 &  & 34.1 & 61.8 & 71.7 & 167.6 \\
CLIP4CLIP [\emph{VaTeX}] & 56.0 & 82.5 & 88.9 & 227.4 &  & 34.0 & 60.3 & 70.6 & 164.9 \\
X-CLIP [\emph{ActivityNet}] & 55.0 & 81.3 & 88.6 & 224.9 &  & 35.6 & 62.5 & 71.2 & 169.3 \\
X-CLIP [\emph{VaTeX}] & 56.9 & \textbf{83.3} & 89.5 & 229.7 &  & 35.0 & 61.0 & 71.0 & 167.0 \\
CLIP-L/14@336px & \textbf{59.5} & 83.2 & \textbf{90.5} & \textbf{233.2} &  & \textbf{45.2} & \textbf{69.4} & \textbf{78.8} & \textbf{193.4} \\
\midrule
\multicolumn{10}{@{}l}{\textbf{\textit{Chinese models:}}} \\
Taiyi & 48.4 & 77.5 & 85.8 & 211.7 &  & 41.1 & 68.2 & 79.8 & 189.2 \\
CN-CLIP & 48.2 & 77.6 & 87.5 & 213.2 &  & 43.8 & 72.2 & 80.5 & 196.4 \\
ERNIE-ViL2 & 53.6 & 81.8 & 89.4 & 224.7 &  & 46.1 & 72.5 & 80.8 & 199.4 \\
CN-CLIP [\emph{VaTeX-CN}] & 59.3     & \textbf{87.2}   & 92.2  & 238.7  &  & 42.0    & 70.1   & 78.3    & 190.4  \\
CN-CLIP [\emph{ChinaOpen-50k}] & \textbf{62.5} & 86.4 &  \textbf{92.7} &\textbf{241.5} &  & \textbf{52.2} & \textbf{79.2} & \textbf{88.0} & \textbf{219.4}\\
\bottomrule
\end{tabular}
}
\end{table}

\subsection{Task III: Video Captioning}

\subsubsection{Task setup}

Video captioning is to generate a natural-language sentence for video content description. Similar to the text-to-video retrieval task, we also setup two tracks. The content-based track uses the manually-written captions as ground truth, while the beyond-content track uses the user-generated title.

For the four image captioning models, \ie mPLUG, BLIP, BLIP-2, and OFA-Chinese, we use the middle frame as their visual input. As for Flamingo, we follow the original paper \cite{alayrac2022flamingo}, uniformly sampling 8 frames per video as its visual input.

As GVT is derived from GIT, the latter is a direct baseline to the former. We thus fine-tune both models on ChinaOpen-50k and VaTeX-CN, respectively. To validate the necessity of our proposed data cleaning pipeline, we randomly selected an equivalent-sized dataset from the raw data, referred to as Bilibili-50k. Moreover, we try to implement a multi-task version of GVT by adding a token-based binary classification head to predict if a given video-text pair is relevant. The head takes as input the [EOS] token, which has seen all preceding tokens. With prompt-based label embedding, open-set video tagging can be performed. Given the caption generation loss ($\ell_g$) and the video-text-matching loss ($\ell_m$), a combined loss is formed as $w\times \ell_g + (1-w) \times  \ell_m$, with weight $w \in [0,1]$.

\subsubsection{Evaluation criteria}

We adopt three common  metrics, \ie  BLEU4, METEOR, and CIDEr, with their mean for overall comparison. The calculation related to the Chinese captions is conducted at a word-level, using Jieba\footnote{\url{https://github.com/fxsjy/jieba}} for Chinese word segmentation.

\subsubsection{Results}

As Table \ref{tab:eval_caption_english_chinese} shows,  with the ability to utilize a pre-trained LLM in a frozen-weights manner, BLIP-2 clearly outperforms the other English models. In the content-based track, the better performance of GIT [\emph{ChinaOpen-50k}] against GIT [\emph{VaTeX-CN}] (42.2 \emph{versus} 23.1) and that of GVT [\emph{ChinaOpen-50k}] against GVT [\emph{VaTeX-CN}] (44.4 \emph{versus} 31.3) shows that ChinaOpen-50k leads to better Chinese video captioning models. The same conclusion can also be drawn from the beyond-content track. Given that ChinaOpen-50k is auto-constructed, these results are encouraging. 

\begin{table}[!ht]
\caption{Performance of video captioning.}
\label{tab:eval_caption_english_chinese}
\centering
 \scalebox{0.65}{
\begin{tabular}{@{}lrrrrrrrrr@{}}
\toprule

\multicolumn{1}{@{}l}{\multirow{2}{*}{\textbf{Model}}}  & \multicolumn{4}{c}{\textbf{Content-based track}} & & \multicolumn{4}{c}{\textbf{Beyond-content track}} \\
 \cmidrule{2-5}  \cmidrule{7-10}  
\multicolumn{1}{c}{} & \multicolumn{1}{l}{BLEU4} & \multicolumn{1}{l}{METEOR} & \multicolumn{1}{l}{CIDEr} & \multicolumn{1}{l}{Mean} & &\multicolumn{1}{l}{BLEU4} & \multicolumn{1}{l}{METEOR} & \multicolumn{1}{l}{CIDEr} & {Mean} \\
\hline
\multicolumn{10}{@{}l}{\textbf{\textit{English models:}}} \\
Flamingo & 8.7 & 7.9 & 27.3 & 14.6 &  & 3.2 & 3.5 & 8.8 & 5.2 \\
mPLUG & 10.9 & 12.8 & 33.7 & 19.1 &  & \textbf{3.9} & \textbf{4.6} & 10.0 & 6.2 \\
GIT & 9.7 & 9.2 & 43.0 & 20.6 &  & 2.2 & 3.1 & 9.2 & 4.8 \\
BLIP  & 17.9 & 13.3 & 62.0 & 31.1 &  & 2.8 & 3.5 & 10.0 & 5.4 \\
BLIP-2 & \textbf{19.5} & \textbf{15.2} & \textbf{79.3} & \textbf{38.0} &  & 3.6 & 4.2 & \textbf{14.3} & \textbf{7.4} \\
\multicolumn{10}{@{}l}{\textit{\textbf{Chinese models:}}} \\
OFA-Chinese & 3.8 & 6.3 & 13.6 & 7.9 &  & 1.1 & 3.0 & 4.3 & 2.8 \\
GIT[\emph{VaTeX-CN}] & 10.7 & 18.4 & 40.1 & 23.1 &  & \textbf{1.7} & 4.3 & 4.9 & 3.6 \\
GVT[\emph{VaTeX-CN}] & \textbf{18.5} & 18.4 & 56.9 & 31.3 &  & 1.6 & 4.5 & 4.8 & 3.6 \\
GIT[\emph{Bilibili-50k}] & 14.9 & 18.4 & 67.8 & 33.7 &  & 0.9 & 3.9 & 6.5 & 3.8 \\
GIT[\emph{ChinaOpen-50k}] & 17.0 & \textbf{19.1} & 90.1 & 42.1 &  & 1.2 & 4.5 & 9.2 & 5.0 \\
GVT[\emph{ChinaOpen-50k}] & 17.7 & \textbf{19.1} &96.3 & 44.4 &  & 1.5 & 4.6 & 9.1 & 5.1 \\
\bottomrule
\end{tabular}
}

\end{table}

\begin{table*}[!b]
\caption{Performance of multi-task GVT on video tagging, retrieval and captioning.}
\label{tab:eval_mt_gvt}
 \scalebox{0.6}{
\begin{tabular}{@{}crrrrrlrrrrrrrrrlrrrrlrrrr@{}}
\toprule
\multirow{4}{*}{} & \multicolumn{5}{c}{\multirow{2}{*}{\textbf{Open-set video tagging}}} &  & \multicolumn{9}{c}{\textbf{Text-to-video retrieval}} &  & \multicolumn{9}{c}{\textbf{Video captioning}} \\
 \cmidrule{8-16}   \cmidrule{18-26} 
 & \multicolumn{5}{c}{} &  & \multicolumn{4}{c}{Content-based track} & \multicolumn{1}{l}{} & \multicolumn{4}{c}{Beyond-content track} &  & \multicolumn{4}{c}{Content-based track} &  & \multicolumn{4}{c}{Beyond-content track} \\
 \cmidrule{2-6}  \cmidrule{8-11}  \cmidrule{13-16}   \cmidrule{18-21}  \cmidrule{23-26} 
 
 $w$ & \multicolumn{1}{l}{Object} & \multicolumn{1}{l}{Action} & \multicolumn{1}{l}{Scene} & \multicolumn{1}{l}{User-tags} & \multicolumn{1}{l}{Mean} &  & \multicolumn{1}{l}{R@1} & \multicolumn{1}{l}{R@5} & \multicolumn{1}{l}{R@10} & \multicolumn{1}{l}{SumR} & \multicolumn{1}{l}{} & \multicolumn{1}{l}{R@1} & \multicolumn{1}{l}{R@5} & \multicolumn{1}{l}{R@10} & \multicolumn{1}{l}{SumR} &  & \multicolumn{1}{l}{BLEU4} & \multicolumn{1}{l}{METEOR} & \multicolumn{1}{l}{CIDEr} & \multicolumn{1}{l}{Mean} &  & \multicolumn{1}{l}{BLEU4} & \multicolumn{1}{l}{METEOR} & \multicolumn{1}{l}{CIDEr} & \multicolumn{1}{l}{Mean} \\
 \cmidrule{2-6}  \cmidrule{8-11}  \cmidrule{12-16}   \cmidrule{18-21}  \cmidrule{22-26} 
 
\multicolumn{1}{r}{0.8} & 16.0 & 17.1 & 17.1 & 11.7 & 15.5 &  & 35.3 & 65.6 & 79.9 & 180.8 &  & 15.8 & 41.4 & 54.9 & 112.1 &  & \textbf{16.6} & \textbf{18.8} & \textbf{87.5} & \textbf{41.0} &  & \textbf{1.5} & \textbf{4.5} & \textbf{9.1} & \textbf{5.0} \\
\multicolumn{1}{r}{0.5} & 16.8 & 18.5 & \textbf{20.8} & 12.8 & 17.2 &  & 39.0 & 70.3 & 81.9 & 191.2 &  & 19.4 & 47.3 & 61.3 & 128.0 &  & 16.2 & 18.5 & 83.9 & 39.5 &  & 1.4 & 4.4 & 8.8 & 4.9 \\
\multicolumn{1}{r}{0.2} & \textbf{20.0} & \textbf{21.6} & 19.3 & \textbf{16.1} & \textbf{19.3} &  & \textbf{40.2} & \textbf{71.3} & \textbf{83.3} & \textbf{194.8} &  & \textbf{22.3} & \textbf{51.6} & \textbf{64.0} & \textbf{137.9} &  & 15.6 & 18.4 & 77.6 & 37.2 &  & 1.2 & 4.2 & 8.4 & 4.6\\
\bottomrule
\end{tabular}
}
\end{table*}

GIT [\emph{ChinaOpen-50k}] is better than its counterpart trained on Bilibili-50k, 42.1 \emph{versus} 33.7 in the content-based track and 5.0 \emph{versus} 3.8 in the beyond-content track. The result justifies the necessity of data cleaning. The performance of multi-task GVT is shown in Table \ref{tab:eval_mt_gvt}. For tagging / retrieval, we see a clear performance gap between the multi-task GVT and the SOTA. Developing a unified model is nontrivial, necessitating further investigation.

\begin{figure}[htb!]
    \centering
    \includegraphics[width=0.97\columnwidth]{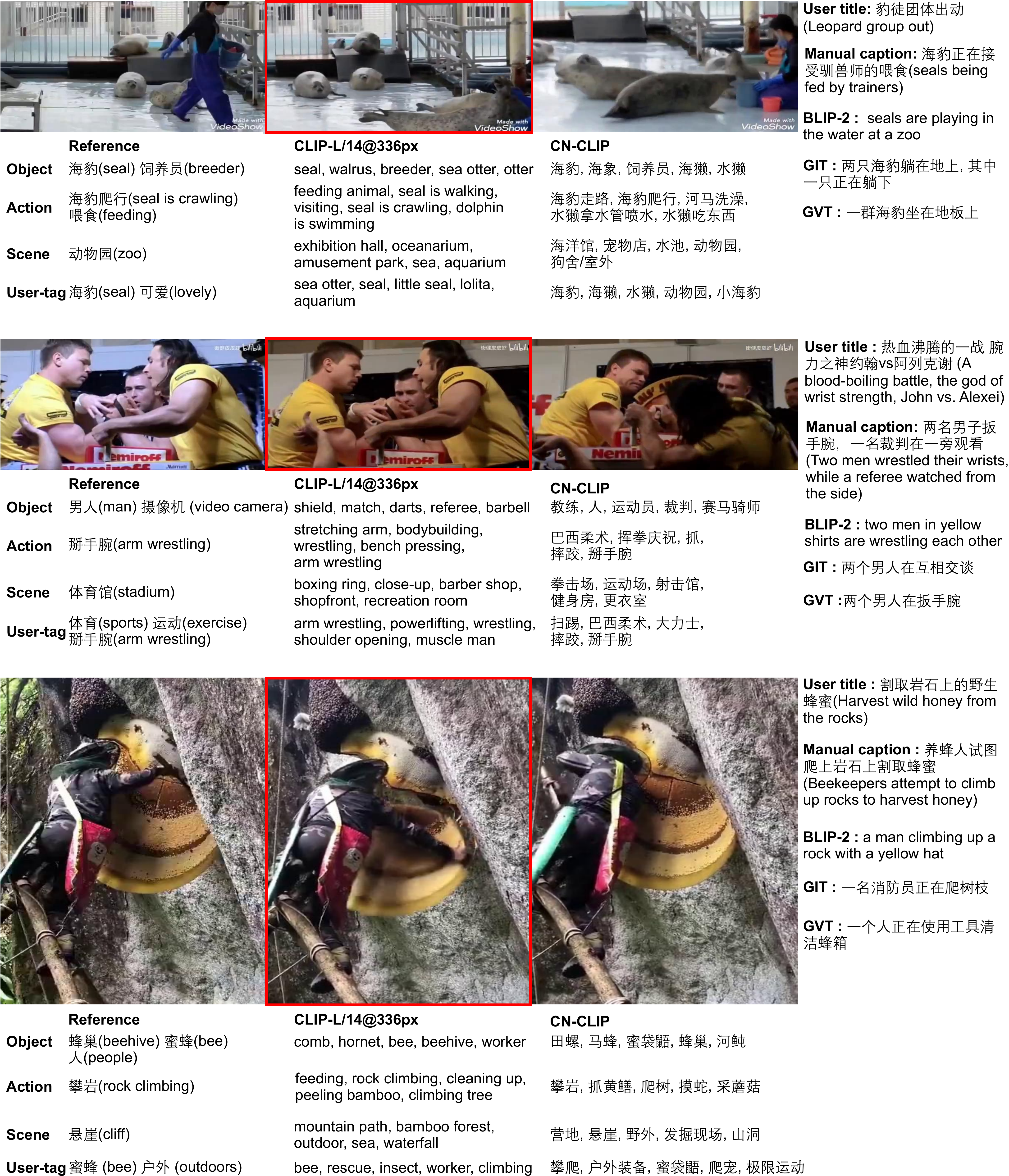}
    \caption{Qualitative results by leading models, \ie CLIP-L/14@336px and CN-CLIP for tagging and BLIP-2, GIT and GVT for captioning. Middle frames are with red borders.}
    \label{fig:demo}
\end{figure}

For both English and Chinese models, their performance scores in the beyond-content track are much lower than their content-based counterparts. Clearly, there is a big gap between what the SOTA video captioning models can generate and what a real user wants his or her videos to be titled. See Fig. \ref{fig:demo} for qualitative results.

\section{Summary and Concluding Remarks} \label{sec:conc}

We develop ChinaOpen, a new video dataset for open-world multimodal learning. The dataset consists of ChinaOpen-50k, a webly annotated video set for training, and ChinaOpen-1k, a manually annotated bilingual video set for testing. Fifteen SOTA models and the proposed GVT  have been evaluated, leading to conclusions as follows. For open-set video tagging, the best English / Chinese model is CLIP-L/14@336px / CN-CLIP. Predicting user tags is more difficult than recognizing objects, actions and scenes. For text-to-video retrieval, CLIP-L/14@336px is again the best English model, while EARNIE-ViL2 is the winning Chinese model. For video captioning, BLIP-2 generates the best English captions, while GVT trained on ChinaOpen-50k generats the best Chinese captions. For both retrieval and captioning tasks, the beyond-content track appears to be more challenging than the content-based track. ChinaOpen has demonstrated a new opportunity for future research.

\textbf{Acknowledgments}.
This work was supported by NSFC (62172420), Tencent Marketing Solution Rhino-Bird Focused Research Program, and Public Computing Cloud, Renmin University of China. The first author thanks H. Doughty and C. Snoek from UvA for helpful discussion on the topic. 

\balance
\bibliographystyle{ACM-Reference-Format}

\bibliography{main}


\end{document}